# Quantifying plasticity: a network-based framework linking structure to dynamical regimes

Igor Branchi

*Center for Behavioral Sciences and Mental Health, Istituto Superiore di Sanità*
Viale Regina Elena, 299, 00161 Roma, Italy, Tel.: +39 06 4990 2833, E-mail: igor.branchi@iss.it
ORCID: https://orcid.org/0000-0003-4484-3598

**Abstract**
Plasticity is a fundamental property of complex systems, such as the brain or an organism. Yet it typically remains a descriptive concept inferred retrospectively from observed outcomes, such as modifications in activity or morphology. Here, the network-based operationalization of plasticity is further formalized as the ratio between system size and connectivity strength among system elements. Within this framework, system size determines the dimensionality of the accessible state space, while connectivity strength tunes the system's regime. An optimal range of plasticity -- balancing capacity for change and capacity to maintain coherence -- emerges at intermediate connectivity strength. Notably, this balance coincides with the critical regime, which provides a theoretically motivated benchmark that enables a normalized unit of measure, termed *effective plasticity*, and comparisons of adaptive efficacy across diverse systems. Plasticity is thus transformed into a predictive tool that quantifies a system's capacity for change before it occurs. Its validity is supported across disciplines and, in particular, by evidence from psychopathology where it anticipates transitions between mental states. At a mechanistic level, plasticity acts as a structural tuning parameter for criticality, reframing their relationship as causal, with plasticity driving criticality rather than merely accompanying it. Furthermore, this network-based operationalization explains how larger systems can more robustly maintain critical dynamics. Crucially, the proposed perspective distinguishes functional regime shifts from thermodynamic phase changes, identifying plasticity as the system-level regulator that shapes and constrains the dynamic repertoire. This framework is applicable across domains, including ecology, economics, and social systems, and may foster cross-disciplinary integration within complexity science.

## 1 Introduction

The concept of plasticity transcends disciplinary boundaries, as it applies to any complex system—at the neurobiological, cognitive, psychological, social, or ecological level – and its foundational features are formulated to hold across domains and levels of analysis (Bateson et al., 2014; Bonamour et al., 2019; Price and Duman, 2020). Plasticity refers to the capacity of a system to modify or adapt in response to contextual perturbations and challenges and is increasingly acknowledged as a key feature of processes involving change, transformation, and development (Branchi, 2022). Plasticity should be understood from a *for-better-and-for-worse* perspective, as it enables not only changes towards improvement, such as recovery, but also towards worsening, determining vulnerability. Therefore, plasticity is not inherently beneficial or harmful, but its value depends on the direction of the changes it enables (Branchi, 2011; Ellis et al., 2011). In the neuroscience domain, relevant studies have described the biological mechanisms that regulate gates on plasticity. For instance, the role of key molecules such as brain-derived neurotrophic factor (BDNF) in underlying synaptic and circuit plasticity has been extensively documented (Castren and Monteggia, 2021). In addition, parvalbumin-positive interneuron maturation and perineuronal nets have been implicated in constraining critical-period plasticity (Hensch, 2004). More recently, the role of glial cells, including microglia (Tremblay et al., 2011) and astrocytes (Sancho et al., 2026), in circuit refinement and stabilization has emerged as a major determinant of adult plasticity.

Building on long-standing theory and evidence that patterns and strengths of interactions constrain a system's dynamical regime and state (Betzel et al., 2023; Holland, 1995; Kauffman, 1993; Parisi, 2006; Sporns, 2022), this article moves from considering connectivity as a descriptor of regime to using it as a defining variable: connectivity and system size are treated as system properties from which plasticity can be defined *a priori* as a system-level, physically grounded quantity. Crucially, the novelty of this framework lies not in the mathematical formalism of critical regime shifts, in the sense of transitions between dynamical regimes, but in identifying plasticity as the specific system level property that emerges from and is determined by network constraints. In this view, plasticity represents the system's dynamic potential within the limits defined by size and connectivity, distinct from their modifications (e.g., rewiring) that tune the plasticity level.

Though the focus of this article is primarily on plasticity in neuroscience and mental health, where it is defined as the capacity to modify brain functioning and mental states (Branchi and Giuliani, 2021), examples from a wide range of fields will be provided, adopting an interdisciplinary perspective. Within the scope of this article, the terms *plasticity* and *flexibility* are used interchangeably, as both refer to the system's capacity for change.

## 2 Two complementary dimensions of plasticity in a network-based operationalization

Classically, plasticity is investigated using experimental paradigms that quantify changes induced by stimulation, experience, or environmental manipulation. In this sense, plasticity is typically inferred retrospectively, based on its observable outcomes. In the neuroscientific and psychological fields, these outcomes correspond to measurable changes in brain structure or function. Notable examples include the seminal work of Bliss and Lømo on long-term potentiation (Bliss and Lømo, 1973), the studies by Hubel and Wiesel on experience-dependent visual cortex development (Wiesel and Hubel, 1963), Rosenzweig's investigations on environmental enrichment and brain morphology (Rosenzweig et al., 1962) and neuroimaging studies reporting experience-related changes in human brain structure or function (Draganski et al., 2004; Maguire et al., 2000).

Building on the network approach to mental health (Borsboom, 2017; Borsboom and Cramer, 2013; Scheffer et al., 2024; Scheffer et al., 2012), a network-based operationalization of plasticity has recently been proposed (Branchi, 2022, 2023). Within this framework, a complex system composed of multiple interacting elements – such as functionally distinct brain regions or the symptoms of a mental disorder – is represented as a weighted graph. Nodes correspond to the system elements and edges represent the connections among them (e.g., correlations of activation patterns), with edge weight quantifying the strength of these connections (Fig. 1).

### *2.1 Transition plasticity*

Within this network representation, the strength of connectivity among the system elements determines the system capacity to change, and therefore constitutes a core dimension of plasticity. This conceptualization is grounded on the idea that, in a highly connected network -- where the behavior of each element is highly correlated with that of the others -- the potential for change of each element is constrained by the need for concurrent modifications across all the interconnected elements. In contrast, in a weakly connected network, individual elements can change more independently as they have fewer or no constraints imposed by the rest of the system (Fig. 1, Top).

Accordingly, this dimension of plasticity, termed *transition plasticity*, has been operationalized as inversely proportional to connectivity strength:

$$Plasticity\ (P) = \frac{1}{\sum_{e \in E} |w(e)|}$$

where $E$ is the set of edges ($e$), $w$ is the edge weight, and $\sum_{e \in E} |w(e)| > 0$ because, in a system, the parts must be at least minimally connected. Based on this formula, plasticity increases as the average strength of the connections decreases and vice versa. In other words, systems characterized by weaker connectivity among their elements exhibit greater transition plasticity. Recent work has emphasized that network structure constrains the dynamical regimes accessible to a system (Betzel et al., 2023; Sporns, 2022). Here, this general principle is leveraged to move beyond a descriptive use of connectivity, treating it as a structural system property that sets the boundaries of the accessible state space and therefore provides the basis for defining plasticity *a priori* as a system-level quantity.

### *2.2 Configurational plasticity*

A second dimension of plasticity becomes relevant when comparing systems that differ in the number of nodes or when considering the same system as its number of nodes increases or decreases. As plasticity is the capacity for change and, thus, for acquiring new configurations in response to contextual inputs, it is inherently related to the number of nodes available for reconfiguration, which determines the dimensionality of the accessible state space (Fig. 1, Bottom). Accordingly, combining this second dimension, termed *Configurational plasticity*, that is proportional to the number of nodes, with Transition plasticity, which is inversely proportional to connectivity strength, yields the following formulation:

$$Plasticity\ (P) = \frac{N}{\sum_{e \in E} |w(e)|}$$

where $N$ is the number of nodes in the network. Figure 2 shows that increasing N produces a proportional upward scaling of plasticity across connectivity values, while preserving the same dependence on connectivity strength.

This formulation does not imply a trivial increase in plasticity with system size. Rather, the number of nodes, $N$, defines the dimensionality of the system's embedding space, which scales with the number of configurational degrees of freedom ($C \propto N$). Specifically, $C$ denotes an upper-bound dimensionality (degrees of freedom), not a count of realizable configurations (Fig. 2).

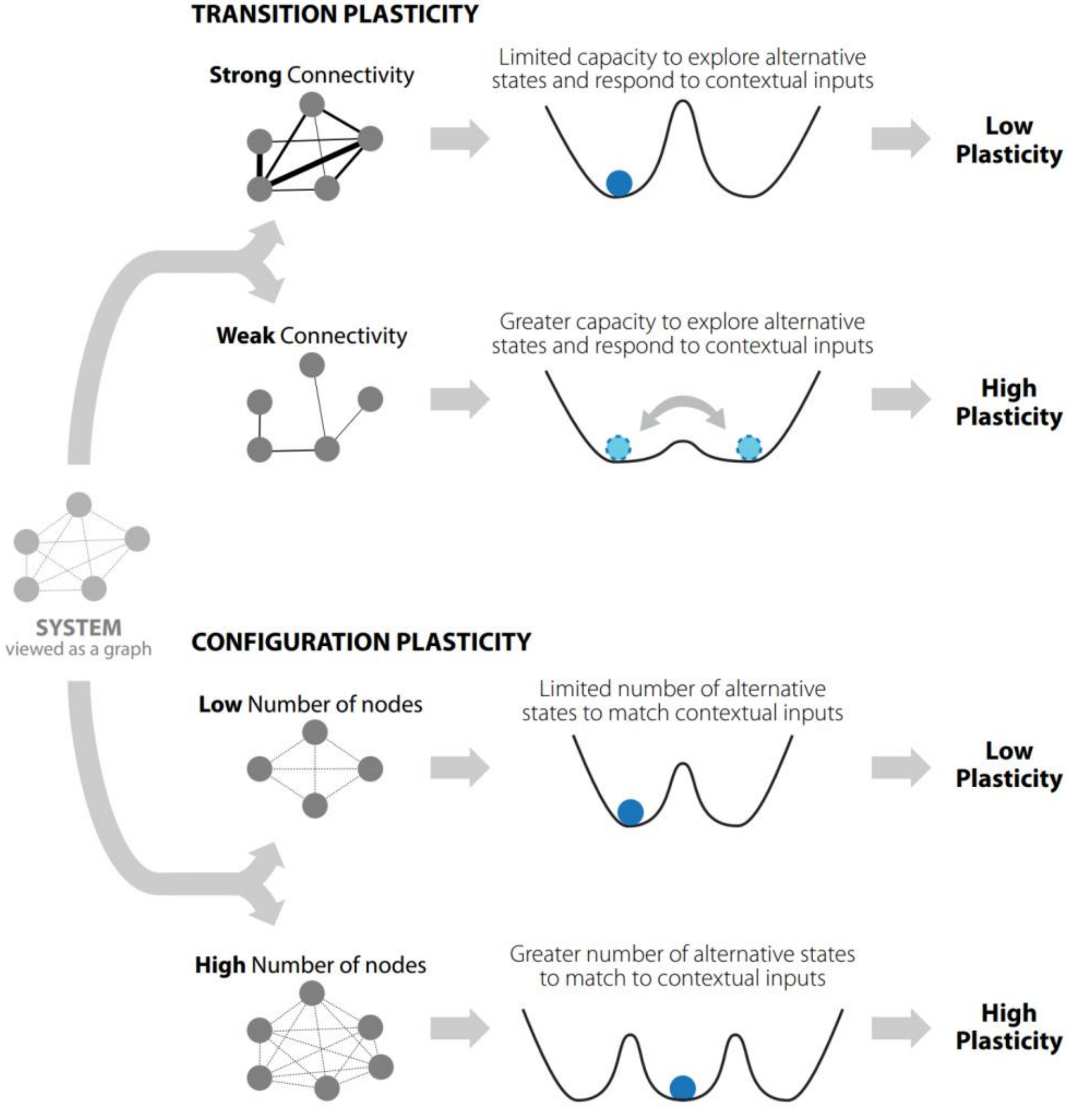

**Figure 1. Two complementary dimensions of the network-based operationalization of plasticity**. When representing a system as a network in which nodes denote system elements and edges denote the connections among them (for example, correlations in their activation), plasticity -- defined as the system's capacity to change and acquire new configurations in response to contextual inputs – is operationalized along two complementary dimensions. **Top**: *Transition plasticity*, which reflects the ease of transitions between states within a fixed state-space dimensionality, is the inverse of connectivity strength. Strong connectivity corresponds to low transition plasticity, represented by deep stability basins that restrict the system state and, therefore, limit the capacity to explore alternative states and respond to contextual inputs. By contrast, weak connectivity flattens the landscape, facilitating transitions between states. **Bottom**: *Configurational plasticity* corresponds to the dimension of plasticity that scales with the number of nodes ($N$). It is proportional to the configurational capacity ($C \propto N$) of the system, as it defines the structural dimensionality of the accessible state space. A larger number of nodes increases the repertoire of alternative configurations available to match contextual demands (represented by a greater number of stability basins). Together, these two dimensions define overall plasticity as the ratio between the number of nodes and connectivity strength. See text for further details.

### 2.3 *Reconciling different views on plasticity*

The proposed framework provides a theoretical bridge for reconciling different views of plasticity in neuroscience and other fields that are often treated as distinct or even contrasting in current literature. On one hand, configurational plasticity aligns with the geometric perspective (Langdon et al., 2023), which focuses on the dimensionality and structure of latent state-space manifolds (i.e., a low-dimensional "space of possibilities" that summarizes high-dimensional state patterns), that is, the latent state space constraining which configurations the system can in principle express. On the other hand, transition plasticity addresses the dynamic perspective (Barack and Krakauer, 2021), which emphasizes the system's ability to navigate its existing repertoire and switch between dynamical states within that latent state space, without necessarily altering its fundamental architecture.

By integrating these components, we show that these perspectives are not mutually exclusive but represent two hierarchical levels of the same phenomenon: configurational plasticity modulates the geometry and extent of the latent state space (i.e., what can, in principle, be represented), while transition plasticity determines accessibility within that space (i.e., what can be reached, and how easily, given current constraints). This integration effectively unifies these approaches. The manifold defines the geometry of the latent state space (i.e., the space of possible system states), while the pattern and strength of interactions among system components (e.g., circuit connectivity in neural systems) provide a key causal constraint that governs transition plasticity and shapes how the system moves through that space.

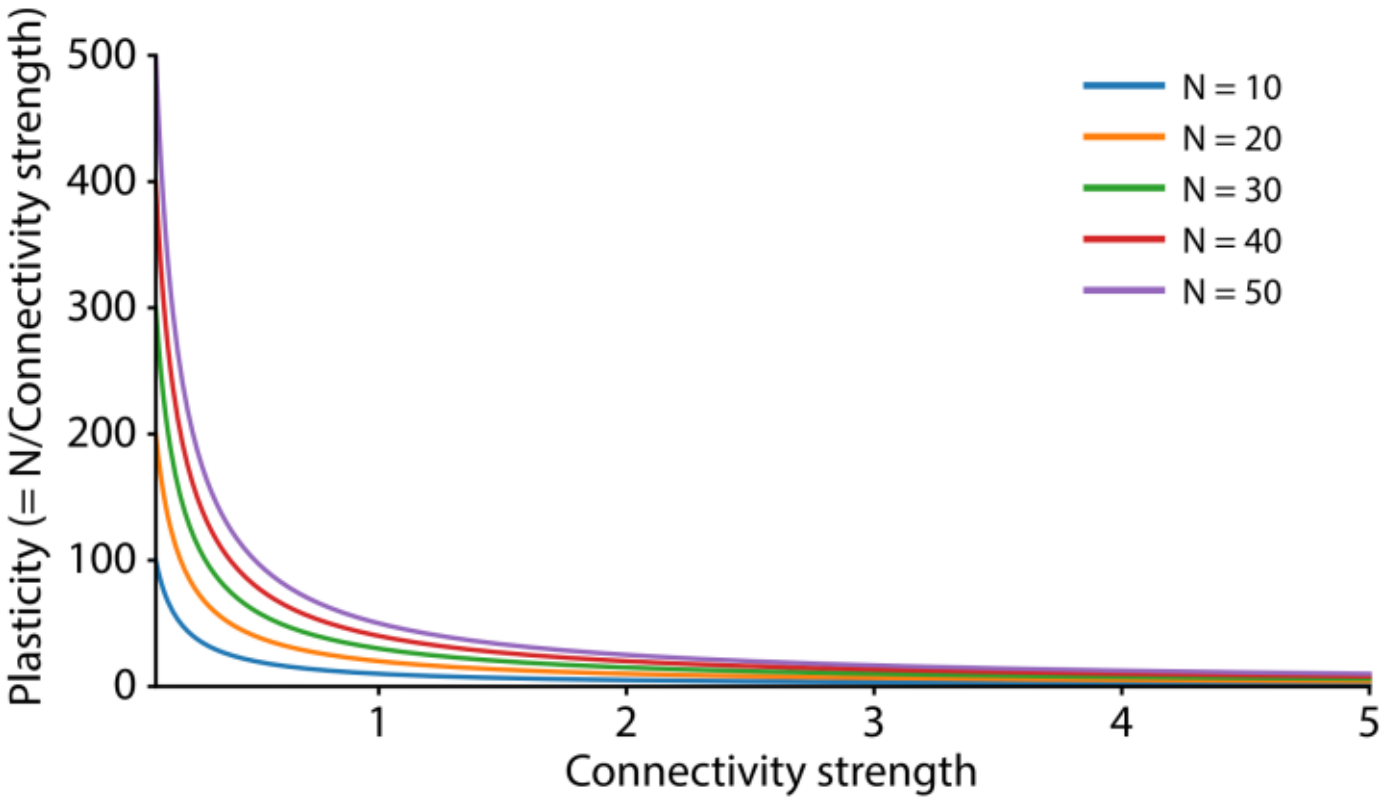

**Figure 2. Plasticity as a function of connectivity strength across system sizes**. Plasticity is defined as the ratio between the number of nodes (N) and connectivity strength. Increasing connectivity strength reduces plasticity in a hyperbolic manner, whereas increasing system size produces a linear upward scaling of plasticity across the entire connectivity range. This representation highlights how system size sets the overall level of plasticity, whereas connectivity strength acts as the primary constraining parameter, regulating the balance between degrees of freedom and coupling within the system.

In this framework, plasticity, which emerges from the interaction between configurational and transition components, captures information that is related to, but not equivalent to, entropic descriptions of state space (Shannon, 1948). While entropy quantifies the distribution of realized states (for example, the entropy of state occupancy across visited configurations), plasticity characterizes the system's capacity to be reshaped by contextual inputs prior to specifying which configurations are most frequently expressed.

### 2.4 *A conceptual framework for complementary dimensions of plasticity*

Overall, *transition plasticity* can be increased by reducing connectivity strength, thereby allowing the system to access a wider range of attainable outcomes without altering the set of available elements. By contrast, *configurational plasticity* is increased by expanding the dimensionality of the system's accessible configuration space through an increase in the number of elements. Together, these two strategies illustrate how plasticity can increase either by weakening constraints among existing elements or by expanding the pool of elements that can be combined, highlighting the complementary roles of connectivity strength and system size. Notably, these two dimensions define the system's dynamic potential for change under a given set of size and connectivity constraints, distinguishing plasticity from structural modifications (e.g., rewiring) that alter these constraints and thereby shift the plasticity level.

Conceptually, the distinction between configurational and transition plasticity resonates with earlier theoretical work on adaptive systems and state-space exploration. For example, Stuart Kauffman's NK models (Kauffman, 1993) and John Holland's framework for complex adaptive systems (Holland, 1995) emphasized how system architecture constrains the exploration of fitness landscapes. However, in these approaches, the capacity to explore configurations is typically inferred retrospectively, based on simulated or observed trajectories across state space. More recently, network neuroscience has highlighted how connectivity constrains accessible dynamical regimes (Betzel et al., 2023). While these contributions compellingly demonstrate that structure matters for dynamics, they do not provide an explicit, prospective, system-level quantity that directly indexes the capacity for change. The present framework extends these insights by offering a unified and predictive operationalization of plasticity grounded in structural network properties, rather than in observed state transitions.

### *2.5 Defining plasticity as an a priori system-level quantity*

The need for such a measure is particularly evident when moving from theoretical landscapes to empirical observations. In contrast to classical approaches described above, such as long-term potentiation or fMRI-based paradigms, which infer plasticity retrospectively from observed outcomes, the present network-based operationalization provides a prospective estimate with predictive value by characterizing a system's capacity for change prior to its empirical manifestation. Therefore, within the framework of complexity science, plasticity can be considered a system-level intrinsic property, and its assessment is explicitly proposed as a derived, physically interpretable system-level quantity that can be quantified and operationalized to describe the system's capacity for change.

The predictive validity of this operationalization of plasticity has been demonstrated in the field of mental health, where higher levels of plasticity have been shown to be associated with an increased likelihood of transitioning across mental states, such as from major depressive disorder to well-being (Delli Colli et al., 2024; Delli Colli et al., 2025a; Delli Colli et al., 2025b). Consistent with this theoretical prediction, distinct plasticity profiles have also been reported in major depressive and bipolar disorders, with the latter exhibiting higher plasticity, aligning with the characteristic fluctuations across diverse psychological states shown by bipolar patients (Viglione et al., 2025). Moreover, the operationalization of plasticity aligns with findings from diverse domains. For example, weak network connections have been shown to facilitate adaptation in both ecological (McCann, 2000; Scheffer et al., 2001) and social systems (Granovetter, 1973; Rajkumar et al., 2022). Similarly, a weakly interconnected financial system tends to exhibit greater adaptability to external shocks (Battiston et al., 2012). Notably, these findings emerged independently and were not originally interpreted within the theoretical framework proposed here.

### *2.6 Applicability across systems and scales*

The present framework is intended to characterize system-level plasticity in finite, interacting systems, under conditions in which global connectivity statistics provide a meaningful description of collective dynamics. While plasticity is conceived here as a general property of complex systems, its quantification requires specifying the level of description at which interactions act as effective constraints on the dynamics of the system. Its applicability therefore relies on several assumptions that delineate its scope. First, the formulation adopts a mean-field perspective, whereby plasticity is defined in terms of aggregate connectivity strength; as such, systems whose dynamics are dominated by extreme topological heterogeneity or by a small number of hubs may require extensions beyond the global average description. Second, the framework assumes a separation of timescales, such that network structure evolves more slowly than the fast dynamics unfolding on it, allowing plasticity to be treated as a structural tuning parameter rather than as a rapidly fluctuating variable (Simon, 1962). Third, the transitions described here are functional regime shifts within a fixed architecture, rather than thermodynamic phase transitions in the renormalization-group sense. Accordingly, they do not require assumptions of universality or the infinite-size limit (Strogatz, 1994). Fourth, connectivity is treated as an effective, scale-dependent quantity capturing functional constraints among elements, rather than as a microscopic interaction parameter. Finally, comparisons across systems require that network nodes represent functionally homologous units, as the choice of elementary units directly affects the quantification of plasticity. Within these conditions, the framework provides a prospective, system-level measure of the capacity for change, while remaining agnostic with respect to the specific direction or qualitative outcome of such change, which is determined by contextual factors.

## 3 Levels of plasticity and the capacity to achieve a novel stable state

Exploring the level of plasticity of a system through its network-based operationalization elucidates the nonlinear relationship between the system's plasticity and its capacity to transition to a stable state (Fig. 3a). A strongly connected system lacks plasticity -- and thus the capacity to transition -- because its tight connections constrain the system's dynamics, limiting exploration of alternative states within the accessible configuration space. As connectivity weakens -- and plasticity increases -- the elements gain the ability to modify independently of one another. This enables the system to reach a regime of *optimal plasticity* in which it can transition across multiple stable states. However, when plasticity exceeds a critical threshold, connections become too weak to sustain stability. This condition, referred to as *instability*, confers a high capacity for change, but disrupts the balance between plasticity and stability, resulting in a system unable to settle into any stable state, despite continued exploration of many transient states within the accessible configuration space. This condition ultimately makes the system unable to transition to a stable state -- much like in the case of strong connectivity, where plasticity is nearly absent. Consequently, the relationship between system plasticity and its capacity to achieve stable states follows an inverted U-shaped curve.

The imbalance between stability and plasticity -- referred to as the stability–plasticity dilemma (McCann, 2000) -- has been previously investigated in multiple scientific domains. For instance, in ecology, stability–plasticity trade-offs have been studied in the context of species adaptation to environmental change, where excessive stability can lead to extinction under shifting conditions, while excessive plasticity may undermine long-term fitness by eroding specialized traits (Chevin et al., 2010; McCann, 2000; Miner et al., 2005). In computational neuroscience and artificial intelligence, models have been developed to address the trade-off through dual-system architectures, such as fast hippocampal learning versus slow cortical consolidation, to preserve both adaptability and coherence (Kumaran et al., 2016; McClelland et al., 1995).

Overall, the relationship between plasticity and stability defines a continuum along which systems exhibit three main functional conditions – rigidity, optimal plasticity, and instability. Among these, only the optimal plasticity regime enables the system to transition and maintain a stable state, thereby supporting effective transitions (Fig. 3a).

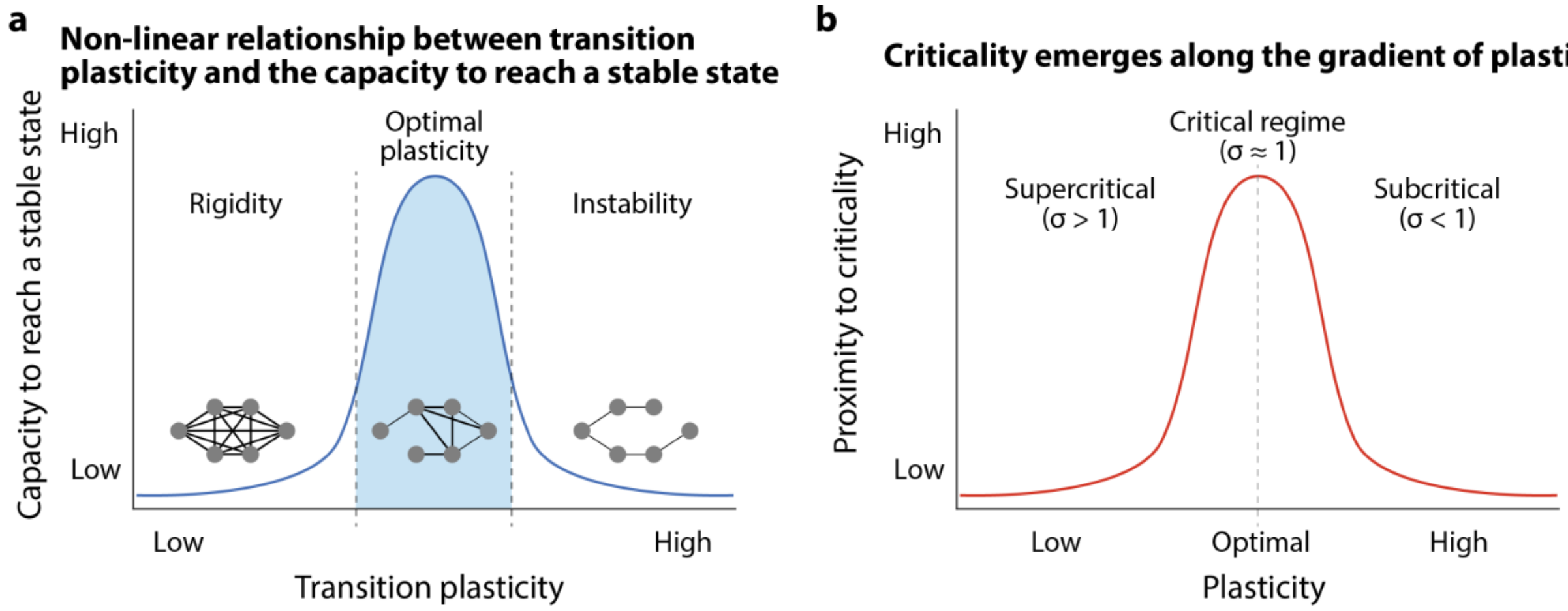

**Figure 3. Plasticity as a dimension linking adaptability and criticality.** (a) The capacity of a system to reach a stable state exhibits a non-linear relationship with plasticity, defined here as the inverse of connectivity strength among system elements, for a given system size. Low plasticity, associated with strong connectivity, results in rigidity and limits the system's ability to transition to novel stable states. Conversely, excessively high plasticity, associated with weak connectivity, leads to instability and prevents the consolidation of new stable states, resulting in high change but low effective adaptation. The capacity to reach a stable state is maximal at intermediate, optimal levels of plasticity, where plasticity and stability are balanced. The shaded blue area indicates this optimal plasticity range. (b) Conceptual illustration of proximity to criticality along the plasticity gradient. The critical regime (corresponding to a branching value $\sigma \approx 1$) coincides with optimal plasticity rather than maximal plasticity, marking the point where the balance between flexibility and stability is achieved. As σ decreases with increasing plasticity, at low plasticity the system is supercritical ($\sigma > 1$), whereas at high plasticity it becomes subcritical ($\sigma < 1$). See text for further details.

## 4 Plasticity and criticality

Criticality refers to the condition in which a system operates at the boundary between order and disorder in a dynamical and functional sense. In this condition, the system exhibits characteristic properties such as scale-free fluctuations, long-range correlations, and an optimal balance between integration and segregation. This balance reflects a trade-off between local specialization and global coordination, a hallmark observed in complex systems including the brain (Bak et al., 1987; Beggs, 2022; Beggs and Plenz, 2003). At the whole system level, these properties expand the repertoire of system configurations, allowing the system to dynamically switch among functional states. Indeed, criticality has been associated with flexibility – which corresponds to plasticity as defined here -- because a system at the critical point is able to effectively adapt to changing environments (Chialvo, 2010; Shew and Plenz, 2013). Although the concept of criticality originated in the study of phase transitions, it is now applied more broadly to describe specific properties of complex systems, from a sandpile to the brain or a flock of starlings (Bak, 1996; Cavagna et al., 2010). In classical statistical physics, criticality is typically achieved by tuning external control parameters, such as temperature or external fields (e.g., Goldenfeld, 1992). By contrast, in the present framework criticality emerges through intrinsic tuning, as plasticity acts as an internal regulator of the system's dynamical regime rather than as an externally imposed control parameter.

The interplay between criticality and plasticity has been previously examined, either through conceptual discussions that link plasticity to the maintenance of critical dynamics or through models showing that synaptic or homeostatic plasticity can tune neural networks toward a critical regime (Beggs and Timme, 2012; Levina et al., 2007; Massobrio et al., 2015; Zierenberg et al., 2018). However, the network-based operationalization of plasticity proposed here introduces a novel analytical and theoretical framework that clarifies how plasticity and criticality converge. Within this framework, and for a given number of nodes, both optimal plasticity and criticality arise when connectivity among system elements is neither excessively strong nor excessively weak. In this intermediate regime, the propagation of signals -- whether expressed as activity, information, or energy -- remains balanced, avoiding both rapid damping and uncontrolled amplification. By contrast, when connectivity among system elements is too strong, corresponding to a supercritical regime, the system is overly rigid and prone to being trapped in specific configurations. Perturbations spread rapidly and uniformly, but this spread is largely saturating and low in information content, so it carries little discriminative information, leading to a loss of diversity and a reduced capacity to explore alternative states. Conversely, when connectivity is too weak, which corresponds to a subcritical regime, the system is unstable. Under these conditions, activity fails to organize coherently, fluctuations become fragmented and transient, and the system struggles to integrate inputs or maintain coordinated behavior. Therefore, both optimal plasticity and critical dynamics emerge only when connectivity among system elements reaches an intermediate level. This condition is not the state of maximal plasticity but rather of optimal plasticity, which reflects the range of connectivity in which the capacity for change and the capacity for retention are balanced. Thus, the relationship between criticality and plasticity follows an inverted U-shaped curve, where both excessively low and excessively high levels of plasticity prevent the emergence of a critical state. Within this framework, criticality corresponds to optimal plasticity, that is, the regime in which the system can both explore novel configurations and reliably stabilize the acquired state (Fig. 3b).

The emergent collective behavior observed in a starling (*Sturnus vulgaris*) flock offers a compelling example to illustrate the non-linear implications of plasticity levels on system criticality and information processing. This phenomenon, characterized by the synchronized aerial maneuvers of thousands of individuals, relies on the continuous adjustment of each starling position in response to local interactions with its nearest neighbors (Cavagna et al., 2010). This adaptive capacity is a direct manifestation of the network connectivity strength -- how much the movement of one starling parallels to that of the others -- and thus systemic plasticity.

When a predator such as a falcon attacks, starlings positioned on the edge of the flock are the first to detect it and react. This reaction propagates like a wave through the entire flock, transmitting information so rapidly that birds on the far side respond even before they can visually perceive the danger. This collective behavior enables the individuals within the flock to evade the falcon as a single body. However, this behavior is effective only if the connectivity among the birds -- that is, the correlation in their flying position -- is close to the critical point, where plasticity is optimal and the balance between plasticity and stability is maintained. If the positions of the starlings are too firmly aligned, the flock becomes overly rigid and enters a supercritical regime, with all individuals locked into a stiff formation. No differential and informative signal can be transferred across the flock because the lack of plasticity hampers the repositioning of one individual relative to the others, hindering not only the propagation of information across the flock but, in cases of extreme rigidity, even its initial acquisition. Conversely, if the positions of the starlings are too loosely coupled, the flock becomes overly plastic and enters a subcritical regime. In this case as well, no information about the approaching predator can be processed by the flock as it quickly dissipates or, in case of extremely high plasticity, it cannot be even acquired: any change in the position of the starling detecting the predator does not stand out from the constant background movements of the other individuals.

### *4.1 Criticality as an emergent property of plasticity levels*

Building on the identification of criticality with optimal plasticity, we next argue that plasticity gradients provide a causal route by which systems reach and maintain the critical regime. The theoretical framework outlined above suggests that criticality emerges along the plasticity gradient, transforming a phenomenological association into a causal relationship. In this view, plasticity is the driver, rather than the byproduct, of criticality. Indeed, the critical point and its characteristic features arise as a consequence of plasticity levels, operationalized as the inverse of connectivity strength for a given number of nodes, and emerge when an optimal balance between plasticity and stability is attained. This provides a mechanistic explanation that may underlie the many foundational observations about the association between plasticity (or flexibility) and criticality (Beggs and Plenz, 2003; Chialvo, 2010; Shew and Plenz, 2013).

Consistently, experimentally validated strategies for shifting systems, including the brain, toward the critical point can be traced back to the network-based operationalization of plasticity, because they all involve modulating the number and/or strength of connections among the system constituent elements. In neuroscience, modulation of synaptic coupling via pharmacological manipulation of excitatory or inhibitory transmission can shift cortical networks toward or away from criticality as shown in organotypic cultures where reducing excessive coupling restored scale-free avalanche dynamics and maximized dynamic range (Shew et al., 2009). Synaptic plasticity mechanisms such as spike-timing-dependent plasticity (Meisel and Gross, 2009) or short-term synaptic depression (Levina et al., 2007) offer additional routes to drive neural networks toward the critical state.

Similar principles extend beyond neuroscience, as criticality has been linked to structural or functional tuning in networks across multiple domains. In genetic regulatory systems, criticality emerges when evolutionary changes in network topology and interaction strengths jointly balance stability and adaptability (Torres-Sosa et al., 2012). In excitable network models, maximal sensitivity and dynamic range occur at the critical point, reached by adjusting the effective branching ratio—that is, the number and strength of active connections (Beggs, 2022; Kinouchi and Copelli, 2006). In infrastructure and communication networks, percolation theory defines a critical threshold of link or node removal at which the giant connected component collapses. The threshold depends directly on the network degree distribution and connection probabilities (Callaway et al., 2000; Cohen et al., 2000). In interdependent networks, catastrophic cascades of failure occur when structural coupling between networks exceeds a critical level (Buldyrev et al., 2010). Across these diverse domains, the evidence consistently converges in demonstrating that criticality reliably emerges from tuning the number, strength, or arrangement of connections, thereby supporting the view that plasticity—operationalized as the inverse of connectivity strength—is the structural driver of criticality.

### *4.2 The links between criticality and plasticity in the light of system size*

The link between criticality and plasticity becomes further evident when considering both the transition and the configurational dimensions by defining plasticity as $P = N / \sum_{e \in E} |w(e)|$. Within this framework, increases in the number of nodes are expected to not only expand the system's degrees of freedom and the space of accessible configurations, but also to enhance the robustness of critical dynamics by reducing the impact of local fluctuations in connectivity on global system behavior through integration across a larger network (Fig. 4). This expansion directly affects scaling relations and fluctuation properties and broadens the dynamic range over which a critical regime can be sustained. As a consequence, larger systems can maintain criticality across wider ranges of coupling strengths or external perturbations, reducing the need for precise fine-tuning. These findings are supported by theoretical and empirical studies showing that system size modulates the manifestation of critical dynamics (Beggs and Plenz, 2003; Beggs and Timme, 2012). In addition, computational models demonstrate that increasing system size expands the state space and that appropriate tuning of connectivity can stabilize dynamics near a critical regime (Levina et al., 2007; Zierenberg et al., 2018). In this view, the explicit dependence on $N$ in the plasticity formulation implies that larger systems can dampen local noise and sustain scale-free dynamics over a broadened range of parameters, reflecting a stabilization rather than a disappearance of the critical zone. Ultimately, the proposed formulation integrates system size as a direct scaling factor and formalizes how the system's capacity for change adapts to increasing system dimensionality, thereby preserving the balance between plasticity and stability that characterizes critical dynamics. Taken together, these results support the interpretation that criticality emerges from a balance between state-space expansion and connectivity constraints, a balance that, within our framework, defines system-level plasticity (see Section 2).

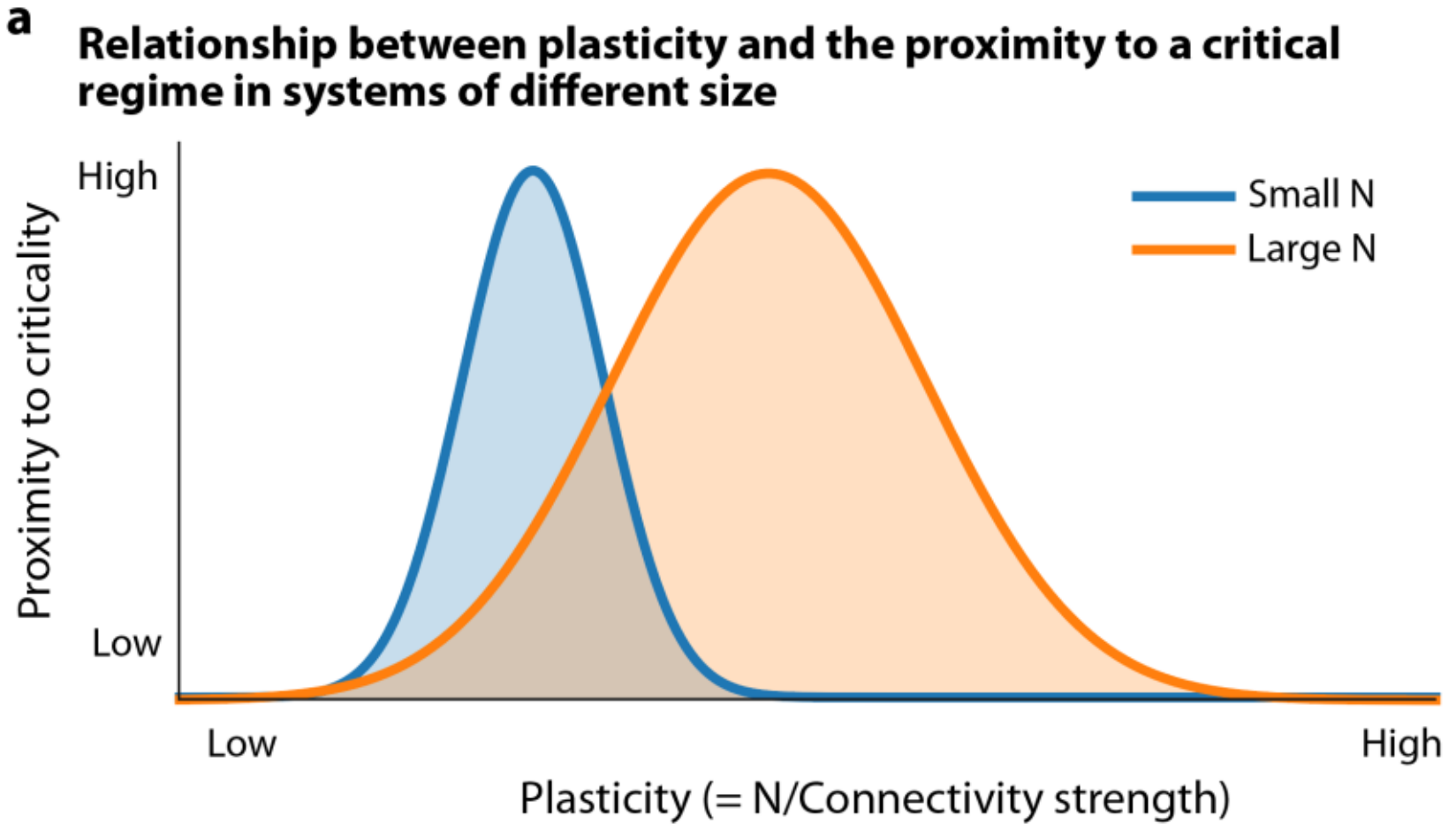

**Figure 4. Effect of system size on plasticity level and criticality range.** Conceptual illustration of the relationship between plasticity -- defined as $P = Number\ of\ nodes\ (N)/\ connectivity\ strength$ -- and the proximity to a critical regime in systems of different size (i.e., with different numbers of nodes). Increasing the number of nodes systematically shifts the system toward higher plasticity values for a given connectivity strength and broadens the functional range of plasticity over which near-critical dynamics can be sustained, reflecting increased tolerance to deviations in connectivity rather than a symmetric change in plasticity itself. Systems with small $N$ exhibit a narrow optimal plasticity range and are highly sensitive to deviations in connectivity, whereas systems with large N maintain near-critical dynamics across a wider range of plasticity values, reflecting increased robustness and a reduced need for fine-tuning. See text for further details.

### *4.3 Reinterpreting and extending Griffiths phases in the light of plasticity*

The framework proposed here naturally incorporates finite-size effects thereby challenging classical thermodynamic assumptions (Weaver, 1948). Unlike structurally homogeneous systems where the critical range collapses to a single point as size increases, heterogeneous networks exhibit extended criticality or Griffiths phases (Moretti and Munoz, 2013; Muñoz, 2018). The operationalization of plasticity takes Griffiths phases into account by formalizing how variations in the connectivity term $\sum_{e \in E} |w(e)|$ reflect the system's internal heterogeneity. Low heterogeneity is reflected by strong overall connectivity, because homogeneous elements respond similarly to inputs and therefore behave uniformly, resulting in highly correlated dynamics. As heterogeneity increases, connectivity weakens, enabling the emergence of locally critical regions and, consequently, an extended critical regime. Mathematically, this reduction in the denominator leads to an increase in plasticity, suggesting that the structural diversity of the network is the driver for the broadening of the critical zone.

Importantly, classical descriptions of Griffiths phases primarily emphasize the emergence and extension of critical-like behavior induced by quenched disorder, without addressing the functional stability of the resulting dynamics (Moretti and Munoz, 2013; Muñoz, 2018). Building on this literature, the present framework introduces an additional stability criterion and predicts that, when heterogeneity is instantiated as excessively weak effective connectivity, the system crosses a functional boundary and enters an overly plastic and unstable regime. In this condition, fluctuations remain spatially confined and the system loses the capacity to consolidate stable states, rendering the extended critical regime functionally unstable (Vojta, 2006;

Zierenberg et al., 2018). Therefore, within this framework, only an intermediate range of heterogeneity supports both extended criticality and functional stability, identifying Griffith-like regimes as a necessary but not sufficient condition for optimal plasticity.

### *4.4 Plasticity as an intrinsic tuning mechanism*

To maintain an optimal balance and avoid the regimes of rigidity or instability described above, the system relies on intrinsic mechanisms to tune the plasticity level. Through ongoing adjustments in connectivity strength, the system explores its parameter space and may reach a functional equilibrium between order and disorder. The framework proposed here aligns with approaches framing psychological and neural processes using thermodynamic and energy-landscape models, emphasizing how entropy, state-space diversity, and dynamic range depend on the underlying organization of interactions (Holland, 1995; Kauffman, 1993; Kringelbach et al., 2024; Parisi, 2006). Within this perspective, structural adaptation operates as the mechanism through which such tuning is implemented at the local level.

This framework finds a strong theoretical foundation in the principles of Self-Organized Criticality (SOC) developed by Per Bak and colleagues (Bak, 1996; Bak et al., 1987). Though differences and potential inconsistencies exist between classical SOC theory and more recent views on criticality -- which often emphasize an input-output matching, high responsiveness, information transmission, or dynamic range -- SOC nonetheless describes, in line with the present framework, how complex systems can spontaneously evolve toward a critical state without the need for an external agent to fine-tune a global control parameter. Specifically, mechanisms of structural adaptation (e.g., rewiring) provide the continuous adjustments in connectivity strength that tune the system's overall plasticity level. Thus, structural modifications determine the constraints, while system plasticity is the resulting dynamic potential. This reframes the relationship between plasticity and criticality from a phenomenological correlation (i.e., critical systems are plastic/flexible) to a causal model in which the system's dynamic potential is the fundamental property that constructs the critical state. Therefore, the hallmarks of criticality, such as long-range correlations, are not merely features of an optimally balanced system, but rather the macroscopic signatures of a system that has successfully self-organized through its inherent plastic dynamics. It is noteworthy that the line of reasoning implies that criticality is a marker of optimal plasticity, defined as an effective balance between plasticity and stability.

In this perspective, plasticity can be interpreted as a measure of dynamically accessible complexity, in that it scales with the structural richness of the system's state space, captured by its size, while being constrained by the strength of interactions that regulate the system's ability to explore that space. Accordingly, plasticity does not merely reflect how complex a system is, but rather how effectively that complexity can be explored and reorganized in response to contextual perturbations. Here, this interpretation is explicitly proposed as an integrative framework, aligning with descriptions of structural richness in the system's state space (Kringelbach et al., 2024; Parisi, 2006), and building on the previously proposed formulation of plasticity that incorporates the constraints imposed by interaction strength (Branchi, 2022, 2023).

## 5 Definition of a unit of measure of plasticity

Despite its central role in adaptive behavior and complex system dynamics across disciplines, from neuroscience to psychology, plasticity is often described as a process and not as a property that can be directly measured. A major advantage of the network-based operationalization proposed here and elsewhere (Branchi, 2023) is that it allows plasticity to be quantified. Specifically, the formula $Plasticity = N / \sum_{e \in E} |w(e)|$ provides a direct quantification, yielding a system-level quantity that reflects the dynamical potential within a system structure.

This unit provides a meaningful absolute description of the system's capacity for change, allowing comparisons of the total configurational repertoire between systems of different sizes (e.g., identifying systems with larger potential state spaces). However, absolute plasticity does not directly indicate the functional state of the system: a large system with high absolute plasticity might still operate far from its optimal regime (e.g., due to excessively strong effective constraints, corresponding to a rigid state), whereas a smaller system might be perfectly tuned. Therefore, to compare adaptive efficacy—defined as how effectively a system utilizes its capacity relative to its structural constraints—a normalized measure is required.

To define this measure, a key observation emerging from the present framework is that optimal plasticity and criticality overlap. Systems exhibit maximal capacity to change while retaining a new stable state when operating near a critical regime, where the balance between plasticity and stability is optimal. Consequently, the branching value ($\sigma$), which indexes critical dynamics, provides a natural, system-independent benchmark. In this framework, $\sigma$ is assumed to vary monotonically along the plasticity gradient, such that systems with low plasticity (strong coupling) are supercritical ($\sigma > 1$), whereas systems with high plasticity (weak coupling) are subcritical ($\sigma < 1$). Specifically, a branching value of unity corresponds to the condition of maximal criticality, and thus defines the condition of optimal plasticity independent of system size or connectivity.

On this basis, effective plasticity, which quantifies the system's proximity to optimal plasticity, is defined as a normalized, dimensionless function bounded between 0 and 1 that indexes proximity to the optimal regime rather than plasticity itself, peaking at the critical branching value ($\sigma = 1$). While the functional form of this curve may vary depending on system-specific constraints (e.g., allowing for asymmetries between transitions toward rigidity versus instability), for the purpose of the present operationalization, we adopt a generalized Lorentzian-like formulation:

$$Effective\ plasticity\ = \frac{1}{\left(1 + \left(\frac{\sigma\ -\ 1}{\gamma}\right)^2\right)}$$

Here, σ denotes the branching value and the parameter γ controls the width of the effective plasticity peak, corresponding to the tolerance of the system to deviations from the optimal critical regime. Crucially, this framework theoretically allows $\gamma$ to be asymmetric ($\gamma_{left} \neq \gamma_{right}$) acknowledging that a system may exhibit different sensitivities to deviations toward supercritical rigidity compared to subcritical instability.

The term effective plasticity is deliberately used to distinguish the measure from the state it indexes. Optimal plasticity refers to a specific regime of system dynamics, corresponding to the condition in which flexibility and stability are maximally balanced. By contrast, effective plasticity quantifies how closely a system operates relative to this regime, reaching its maximum value at criticality and decreasing gradually as the system departs toward either rigidity or instability.

Together, the dimensional plasticity derived from the base formula and the normalized measure of effective plasticity provide complementary descriptions. The former captures absolute plasticity within a system and allows for comparisons of potential magnitude across scales, while the latter establishes a principled unit that enables the adaptive efficacy to be compared across diverse systems.

Importantly, the definition of the system's elementary unit directly influences the quantification of plasticity, as excessive coarse-graining may obscure internal dynamics (Betzel et al., 2023). However, the proposed framework renders this scale dependence explicit and theoretically interpretable. Within this view, configurational plasticity captures the magnitude of structural resources available for change, whereas effective plasticity quantifies the efficiency with which these resources are organized relative to the system's functional optimum.

## 6 Criticality operates through permissive causality and yields context-dependent outcomes

As criticality and plasticity converge conceptually as expressions of a system's capacity for change, the significance of criticality can be interpreted in the light of the functional role of plasticity. For example, when brain activity reorganizes in response to experience to support learning, a supercritical regime would render the system too rigid to implement change, but an optimal plasticity level -- overlapping with the critical regime -- is required for adaptive reorganization (Kringelbach et al., 2024). Conversely, a subcritical regime—marked by weakened constraint on network interactions—may reflect an unstable high-plasticity state that places the nervous system at risk of functional breakdown, as shown in early Alzheimer's disease progression (Javed et al., 2025). However, criticality is not universally the preferred regime. Rather, depending on the levels of capacity for change, a system will operate in a critical or supercritical regime. This occurs, for instance, when the system must preserve evolutionarily consolidated functions despite external perturbations (Tinbergen, 1963).

A further lens through which to understand the function of criticality emerges from the interplay between plasticity and context. As noted above, plasticity is, by definition, not inherently good or bad because it enables change without determining its direction. Accordingly, plasticity operates through permissive causality amplifying the system's capacity to change in response to contextual inputs by increasing both the accessibility and the range of attainable trajectories, without specifying whether these trajectories are adaptive or maladaptive (Branchi, 2011, 2022). The direction of change is determined, at least in part, by the quality of context. This is corroborated by evidence across neuroscience and mental health showing that interventions enhancing plasticity promote changes shaped by contextual factors. For example, in preclinical models, plasticity induced by selective serotonin reuptake inhibitors (SSRIs) produces beneficial effects when paired with favorable conditions and detrimental effects when paired with adverse environments (Alboni et al., 2017; Branchi et al., 2013; Maya Vetencourt et al., 2008). At the clinical level, individuals characterized by high plasticity --as measured through a network-based operationalization --show greater and faster improvement in favorable but not in adverse contexts (Delli Colli et al., 2024;

Delli Colli et al., 2025b). Because criticality aligns with plasticity and the value of the latter is context-dependent, the value of criticality is also context-dependent. Achieving criticality is advantageous when paired with conditions that promote beneficial change. However, when contextual conditions promote detrimental change, a critical regime may increase vulnerability by amplifying the system's capacity to follow maladaptive trajectories. Thus, criticality is not inherently beneficial or detrimental, and its value depends on the quality of the outcomes emerging from the changes it fosters. Accordingly, quantification of the proximity to criticality may inform interventions aimed at modulating cognitive flexibility, learning efficiency, and resilience by targeting plasticity-related network properties (Branchi, 2025a; Kringelbach et al., 2024).

The convergence between plasticity and criticality highlights another potential feature of criticality: it operates through permissive causality, increasing or reducing the repertoire of attainable system states and enabling or constraining transitions among them (Branchi and Giuliani, 2021). However, criticality does not dictate the specific outcome that a system or an individual, such as the brain, will achieve. In contrast, the achieved outcome is determined by factors that act through instructive causality, such as the contextual ones (Branchi, 2024, 2025b). Overall, criticality sets the capacity of a system to transition without setting the direction of such transitions.

## 7 Conclusions

The network-based operationalization of plasticity enhances both the theoretical depth and the applied potential of this construct (Branchi, 2022). By formulating it as inversely related to connectivity strength and proportional to the number of nodes within a network, it becomes possible to quantify a system's capacity for change (Branchi, 2023). Crucially, the present framework extends classical approaches to critical phenomena by distinguishing thermodynamic phase transitions from transitions between functional states occurring within a given regime, and by proposing plasticity as a system-level quantity that governs the range of dynamically accessible configurations. In this view, while plasticity is typically inferred retrospectively as a process or outcome, here it is defined prospectively as an intrinsic property determined by network structure.

More broadly, this framework offers a novel perspective on the emergence of complex system dynamics, in which plasticity is conceived as a mechanism that tunes systems toward criticality. Departures from this regime may underpin both rigidity and instability. In the mental health field, both insufficient and excessive plasticity may therefore compromise the individual's ability to achieve and maintain psychological states, such as mental well-being. The present framework complements computational theories including predictive processing (Friston, 2010), attractor-network models of memory (Deco et al., 2009), and reinforcement-learning models of exploration and exploitation (Gershman and Daw, 2017), by providing a system-level, structure-based characterization of a system's capacity for change, rather than specifying algorithmic or representational mechanisms.

Finally, plasticity emerges as a novel and informative parameter space for complex systems, enabling a richer classification of system states, from rigidity to instability, with optimal plasticity and criticality emerging in the intermediate regime. In this sense, plasticity provides a unifying perspective that may foster cross-disciplinary dialogue across complexity science.

**Acknowledgements**
I am grateful for the comments and inputs received from many colleagues and friends: Alessandro Giuliani, Giorgio Ausiello, Maurizio Mattia, Gianni Valerio Vinci, Silvia Poggini, Aurelia Viglione and Claudia Delli Colli. I thank also Stella Falsini, Flavio Torriani and Antonio Maione for logistic support. The research was supported by the grant from the ERANET Neuron and Istituto Superiore di Sanità, project EnviroMood, and the and Istituto Superiore di Sanità, project NeuroDynamics ISS20-6bbea6c1e02b, both to IB.